\documentclass[preprint,prb,superscriptaddress,amssymb,amsmath, floatfix,citeautoscript]{revtex4-1}
\usepackage[pdftex]{graphicx}
\usepackage{color}

\begin{document}

\title{Fresnel transmission coefficients for thermal phonons at solid interfaces}
\author{Chengyun Hua\textsuperscript{a}}
\author{Xiangwen Chen\footnote{\text{
These authors equally contributed to this work.}}}
\author{Navaneetha K. Ravichandran}
\author{Austin J. Minnich}
\email{aminnich@caltech.edu}
\affiliation{%
 Division of Engineering and Applied Science\\
 California Institute of Technology, Pasadena, California 91125,USA
}%

\date{\today}

\begin{abstract}

Interfaces play an essential role in phonon-mediated heat conduction in solids, impacting applications ranging from thermoelectric waste heat recovery to heat dissipation in electronics. From a microscopic perspective, interfacial phonon transport is described by transmission and reflection coefficients, analogous to the well-known Fresnel coefficients for light. However, these coefficients have never been directly measured, and thermal transport processes at interfaces remain poorly understood despite considerable effort. Here, we report the first measurements of the Fresnel transmission coefficients for thermal phonons at a metal-semiconductor interface using ab-initio phonon transport modeling and a thermal characterization technique, time-domain thermoreflectance. Our measurements show that interfaces act as thermal phonon filters that transmit primarily low frequency phonons, leading to these phonons being the dominant energy carriers across the interface despite the larger density of states of high frequency phonons. Our work realizes the long-standing goal of directly measuring thermal phonon transmission coefficients and demonstrates a general route to study microscopic processes governing interfacial heat conduction. 

\end{abstract}

\maketitle

\clearpage

Transport across interfaces in heterogeneous media is a fundamental physical process that forms the basis for numerous widely used technologies. For example, the reflection and transmission of light at interfaces, as described by the Fresnel equations, enables wave-guiding with fiber-optics and anti-reflection coatings, among many other functionalities. Interfaces also play an essential role in phonon-mediated heat conduction in solids.\cite{Cahill2014Review} Material discontinuities lead to thermal phonon reflections that are manifested on a macroscopic scale as a thermal boundary resistance (TBR), also called Kapitza resistance, $R_k$, that relates the temperature drop at the interface to the heat flux flowing across it. TBR exists at the interface between any dissimilar materials due to differences in phonon states on each side of the interface.\cite{Swartz1989} Typical interfaces often possess defects or roughness which can lead to additional phonon reflections and hence higher TBR. 

TBR plays an increasingly important role in devices, particularly as device sizes decrease below the intrinsic mean free paths (MFPs) of thermal phonons.\cite{Cahill2014Review} At sufficiently small length scales, TBR can dominate the total thermal resistance. For instance, the effective thermal conductivity of a superlattice can be orders of magnitude smaller than that of the constituent materials due to high TBR.\cite{pettersson_theory_1990, Chen1998,Ravichandran2014,Chen2013} This physical effect has been used to realize thermoelectrics with high efficiency\cite{SciencePaper,Biswas2012} and dense solids with exceptionally low thermal conductivity\cite{Chiritescu2007}. On the other hand, TBR can lead to significant thermal management problems\cite{Pop2010,Moore2014,Cho2015} in applications such as LEDs\cite{Su2012,Han2013} and high power electronics\cite{Yan2011,Cho2015}.

Thus both scientifically and for applications, a fundamental understanding of thermal transport across solid interfaces is essential. In principle, Fresnel transmission coefficients can also be used to provide a microscopic description of thermal phonon transport at interfaces owing to the similarities between photons and phonons. However, despite decades of work, the microscopic perspective of heat transport across interfaces remains poorly developed compared to that available for photons. Today, interfaces are most often studied using macroscopic measurements of TBR or thermal conductivity. For example, numerous works have studied interfacial thermal transport by observing the temperature dependence of the thermal conductivity\cite{Wang2011} or interface conductance, $G = 1/R_k$\cite{Lyeo2006,Norris2009,Cheaito2015,Schmidt2010,duda_role_2010} or by correlating changes in bonding strength and interface conductance.\cite{OBrien2012,Losego2012} However, these experimental approaches provide limited information about the transmission coefficients because the observable quantities are averaged over all phonons and thus obscure the microscopic processes of interest. 

As a result, considerable uncertainty exists as to the values of the phonon transmission coefficients at solid interfaces. Commonly used analytical models for the transmission coefficients include the gray model, in which the transmission coefficients are a constant for all phonon modes,\cite{bux_nanostructured_2009,Joshi2008,Minnich2009} the acoustic mismatch model\cite{Khalatnikov1952,Little1959} and the diffuse mismatch model (DMM)\cite{Swartz1987,Hopkins2010,hopkins_effects_2007}, yielding incompatible predictions for the transmission coefficients. Further, their predictions are in poor agreement with experiment, and none are able to account variations in the atomic structure of actual interfaces. Atomistic methods such as molecular dynamics\cite{Maiti1997,Stevens2007,Landry2009,IhChoi2012,Jones2013,Yang2013,Merabia2014,Liang2014,Schelling2002b} and atomistic Green's functions\cite{Zhang2006, Li2012b,Tian2014,Huang2010,Hopkins2009} have been extensively applied to obtain the transmission coefficients, predicting the phonon transmission coefficients to decrease with increasing phonon frequency. However, no direct experimental verifications of these predictions have been reported. Prior experimental observations have reached conflicting conclusions, with one reporting that the transmission coefficients follow the trend of atomistic calculations\cite{Wang2011} while another reaching the opposite conclusion that high frequency phonons are the dominant heat carriers across the interface.\cite{Wilson2014b} Therefore, despite considerable experimental and theoretical study, the phonon transmission coefficients at actual interfaces remain unclear. 

Here, we report the first measurements of the Fresnel transmission coefficients for thermal phonons at a solid interface. Our approach is based on applying our recent advances in ab-initio phonon transport modeling based on the phonon Boltzmann transport equation (BTE) to interpret measurements from the time-domain thermoreflectance (TDTR), a widely-used optical experiment in the thermal sciences. Our measurements reveal that low frequency phonons are nearly completely transmitted across solid boundaries while high frequency phonons are nearly completely reflected, leading to an interfacial heat flux distribution that is dominated by low to mid frequency phonons. Further, our approach demonstrates a general route to directly experimentally study the microscopic transport processes governing interfacial heat conduction.

\section*{TDTR with ab-initio phonon transport modeling}

Our measurement is based on the TDTR experiment, an optical pump-probe technique that is used to characterize thermal properties on micron length scales. In this experiment, a femtosecond pulsed laser beam is split into a pump and a probe beam. The pump pulse train is modulated at a frequency from 1 to 15 MHz to enable lock-in detection, and is then used to impulsively heat a metal film coated on the sample. The transient temperature decay at the surface is detected as a change in optical reflectance by the probe beam.\cite{Capinski1996} Typically, this temperature decay curve is fitted to a standard multilayer heat diffusion model based on Fourier's law with the substrate thermal conductivity and metal-substrate interface conductance as fitting parameters.\cite{Cahill2004,Schmidt2008} This method is now a standard thermal metrology technique. 

\begin{figure*}[t!]
\centering
\includegraphics[scale = 0.25]{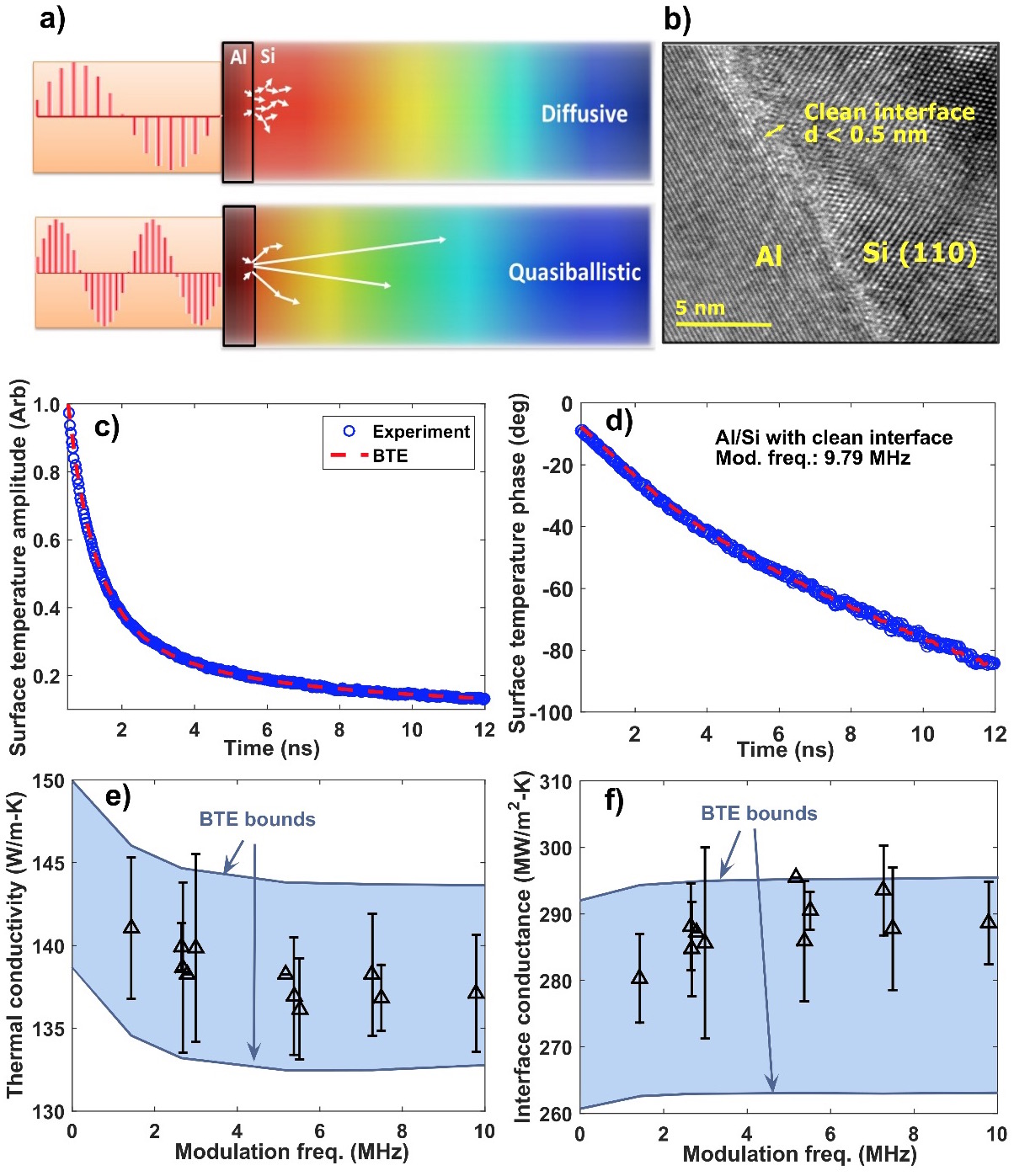}
\caption{\textbf{Measurements and simulations of TDTR experiments on Al/Si with a clean interface.} (a) Schematic of the principle underlying the measurement of transmission coefficients. If the characteristic length scale of the thermal transport is much longer than the phonon MFPs, information about the interfacial distribution is lost due to strong scattering. If some MFPs are comparable to the thermal length scale, the non-equilibrium distribution at the interface propagates into the substrate where it can be detected. (b) TEM image showing the clean interface of a  Al/Si sample with the native oxide removed. The interface thickness is less than $0.5$ nm. Experimental TDTR data (symbols) on this sample at $T = 300$ K for a modulation frequency $f = 10.2$ MHz along with the (c) amplitude and (d) phase fit to the data from the BTE simulations (dashed lines), demonstrating excellent agreement between simulation and experiment. (e) Apparent thermal conductivity and (f) apparent interface conductance of the experiments (symbols) and BTE simulations (lines) versus modulation frequency. These quantities are extracted by fitting the data and simulations to a thermal diffusion model. The pairs of solid lines denoted BTE bounds correspond to the uncertainty in the measured transmission coefficients plotted in Fig.~\ref{fig2}.}
\label{fig1}
\end{figure*}

Recently, considerable interest has focused on quasiballistic heat conduction in TDTR in which a lack of phonon scattering in the substrate leads to the breakdown of Fourier's law. Many experimental reports have demonstrated clear evidence of quasiballistic heat transport in different material systems in the form of thermal properties that appear to deviate from their bulk values.\cite{Koh2007,Siemens2010,Minnich2011a,Regner2012,Johnson2013,Vermeersch2015a} In this work, we interpret these effects as fundamentally originating from the non-equilibrium phonon distribution emerging from the interface. As illustrated in Fig.~\ref{fig1}(a), when MFPs are much shorter than the characteristic length scale of the thermal gradient, information about the phonon distribution at the interface is lost due to scattering. On the other hand, if some phonons have sufficiently long MFPs, the non-equilibrium distribution penetrates into the substrate and affects the resulting heat conduction, thereby providing direct information about the spectral phonon distribution at the interface. 

While the required data is straightforward to obtain, the key to extracting the transmission coefficients is to rigorously interpret the data with the ab-inito phonon transport modeling based on the BTE. It is this step that has long impeded efforts to study interfaces due to the extreme cost of solving the BTE for the TDTR experiment. A number of simplified models\cite{Maznve2011,Minnich2011b,Wilson2013,Vermeersch2015a,Regner2014,Koh2014,Maassen2015} have been proposed to explain these observations. However, all of these models make various approximations that limit the information that can be extracted from the measurements. 

In this work, we overcome this challenge using two recent advances we reported for rigorously solving the spectral BTE under the relaxation time approximation (RTA), with no additional simplifications, that yield a factor of $10^4$ speedup compared to existing methods and allows the first ab-initio phonon transport modeling of TDTR free of artificial parameters or simplifications of the phonon dispersion. First, we have obtained an analytical solution of the spectral BTE in a semi-infinite substrate subject to an arbitrary heating profile.\cite{Hua2014b} Second, we have employed a series expansion method to efficiently solve the spectral, one-dimensional (1D) BTE in a finite layer, suitable for the transducer film.\cite{Hua2014c} In this work, these two solutions are combined using a spectral interface condition\cite{Minnich2011b} that expresses the conservation of heat flux at each phonon frequency. Following the conclusions of prior computational works\cite{Lyeo2006,Murakami2014} that the energy transmission at the interfaces considered here is elastic, we enforce that phonons maintain their frequencies as they transmit through the interface; direct electron-phonon coupling and inelastic scattering is neglected (see Supplementary Information).

The details of the calculation are in the Supplementary Information. The only inputs to our calculation are the phonon dispersions and lifetimes, calculated using density functional theory (DFT) with no adjustable parameters by Jes$\acute{\text{u}}$s Carrete and Natalio Mingo, and the only unknown parameter is the spectral transmission coefficients across the interface. We adjust the spectral transmission coefficients to obtain the best fit of the simulated surface temperature decay to the experimental measurement. The reflection coefficients are specified by energy conservation once the transmission coefficients are known.\cite{Minnich2011b}

\section*{Measurements of phonon transmission coefficients}
We demonstrate our transmission coefficient measurements by performing a TDTR measurement of an Al film on Si substrate with the native oxide removed by Hydrofluoric acid prior to Al deposition, yielding a clean interface. The TEM image in Fig.~\ref{fig1}(b) shows the interface thickness is less than 0.5 nm. The amplitude and phase of a typical signal from the lock-in amplifier are given in Figs.~\ref{fig1}(c) \& (d). To begin, we follow the typical procedure of fitting data with the standard heat diffusion model to extract thermal conductivity $k$ and interface conductance $G$.\cite{kang_two-tint_2008, schmidt_pulse_2008} We obtain $G \approx 280$ MW/m$^2$-K and $k \approx 140$ W/m-K, which is in good agreement with prior works and literature values for the thermal conductivity of Si.\cite{Minnich2011a,Wilson2014b}

Due to the good agreement, this measurement is typically taken as evidence that phonon MFPs in silicon are small compared to the thermal length $\sqrt{\frac{\alpha}{f_0}} \sim1\ \mu$m, where $\alpha$ is the thermal diffusivity of silicon and $f_0$ is the modulation frequency. However, a number of ab-initio calculations, by different groups and with different computational packages, clearly show that a substantial amount of heat is carried by phonons with MFPs exceeding 1 $\mu$m.\cite{Broido2007,Esfarjani2011} This prediction has recently been experimentally confirmed by Cuffe \emph{et al} using thermal measurements on variable thickness silicon membranes.\cite{Cuffe2014} This fact implies that quasiballistic transport should be readily observable in a typical TDTR experiment on Si, despite the seemingly correct thermal properties measured. 

This apparent contradiction can be resolved by considering the spectral profile of the transmission coefficients. Our BTE calculations reveal that the measurements from TDTR on this sample strongly depend on the spectral profile of the transmitted phonon spectrum, a dependence that does not occur in the heat diffusion regime. This dependence allows the transmission coefficients to be directly obtained from the TDTR data by adjusting them to fit the surface temperature data from the lock-in amplifier in a procedure exactly analogous to that used to measure bulk thermal properties.

We performed the fitting of transmission coefficients by adjusting them until the simulated surface temperature curve and experimental data from the lock-in amplifier matched each other. This comparison for all samples can be found in Supplementary Information. To more compactly report the data, we further process the BTE results by fitting the simulated surface temperature decay curves to the same heat diffusion model used in the experiments to extract the apparent thermal conductivity and interface conductance at different modulation frequencies. If these two parameters at each modulation frequency match, then the fitting curves will also match, enabling a compact comparison of the simulation and experimental data sets. However, we emphasize that the use of the heat diffusion model is for ease of comparison only and was not used in the transmission coefficient measurement. 

\begin{figure*}[t!]
\centering
\includegraphics[scale = 0.55]{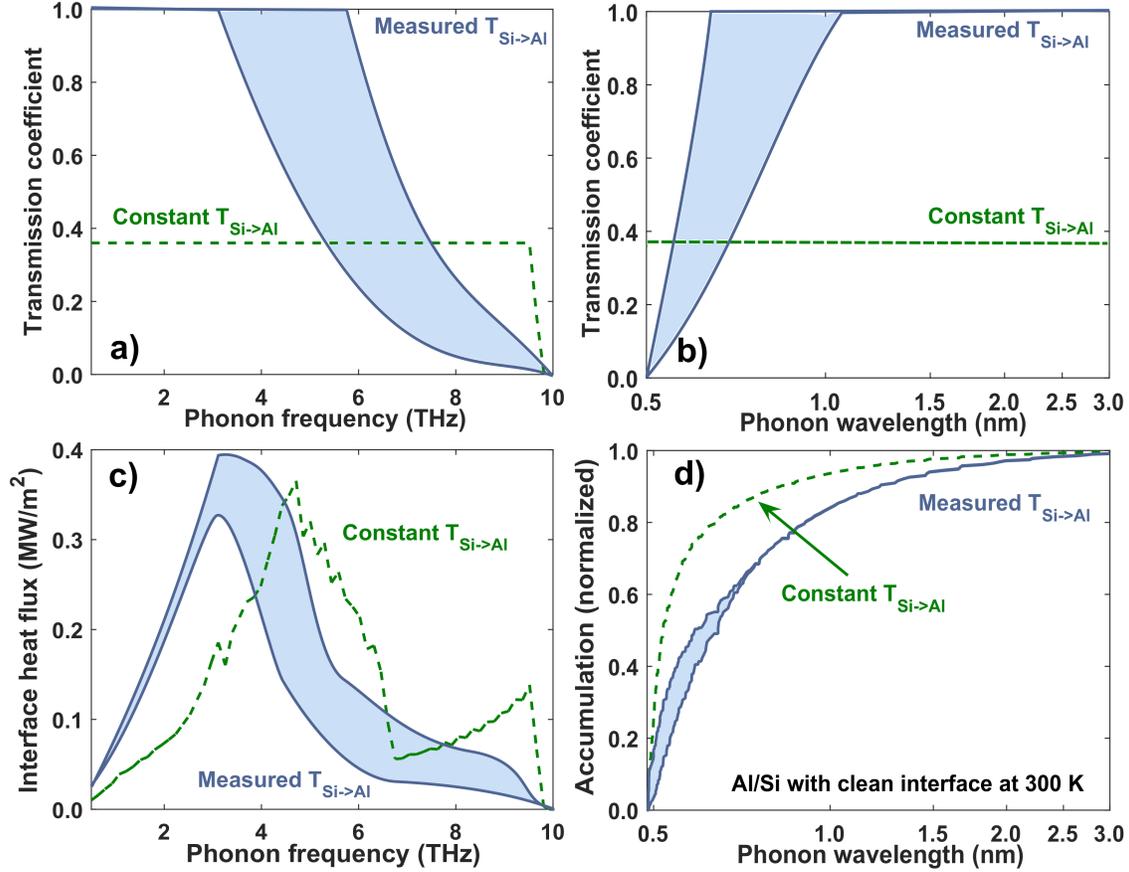}
\caption{\textbf{Transmission coefficients and spectral heat flux at the interface.} Transmission coefficients of longitudinal phonons T$_{\text{Si} \rightarrow \text{Al}}(\omega)$ (blue shaded region) versus (a) phonon frequency and (b) phonon wavelength, along with the constant transmission coefficient profile (green dashed line) that gives the same interface conductance as the measured T$_{\text{Si} \rightarrow \text{Al}}(\omega)$. The boundaries of the shaded region correspond to the BTE bounds in Figs.~\ref{fig1}(e) and (f).  Our measurement shows that low frequency phonons have a much higher transmission probability across the interface than high frequency phonons. (c) Spectral heat flux with the measured (blue shaded region) and constant (green dashed line) transmission coefficient profiles across the interface versus phonon frequency. (d) Normalized accumulative interface conductance with the measured (blue shaded region) and constant (green dashed line) transmission coefficient profile versus phonon wavelength. Contrary to the prediction of the constant transmission coefficient profile, low-frequency/long-wavelength phonons carry a significant amount of heat across the interface.}
\label{fig2}
\end{figure*}

An example of the measurement process for the data shown in Figs.~\ref{fig1}(e) \& (f) is given in Fig.~\ref{fig2}. Prior works\cite{Minnich2011b,Ding2014} used a constant transmission coefficient profile that explained the apparent interface conductance. However, we find that this profile predicts a modulation-frequency dependent apparent thermal conductivity that becomes as low as 100 W/m-K, in strong disagreement with experiment. Similarly, other commonly used models such as the DMM predict the wrong trend of thermal conductivity and interface conductance as a function of modulation frequency. (See Supplementary Information)

The only transmission coefficient profile that is able to simultaneously match the apparent thermal conductivity and interface conductance in Figs.~\ref{fig1}(d) \& (e) from the experiment is shown in Fig.~\ref{fig2}(a). The figure shows that the transmission coefficient from Si to Al for longitudinal phonons, T$_{\text{Si} \rightarrow \text{Al}}(\omega)$, starts at unity, its maximum possible value and decreases steadily to near zero for high phonon frequencies. The transmission coefficient profiles for the other polarizations have similar shapes, and so throughout the paper we plot only the longitudinal transmission coefficients for simplicity. The transmission coefficients from Al to Si, T$_{\text{Al} \rightarrow \text{Si}}(\omega)$ are calculated by satisfying the principle of detailed balance; the relationship between T$_{\text{Si} \rightarrow \text{Al}}(\omega)$ and T$_{\text{Al} \rightarrow \text{Si}}(\omega)$ reflect the differences in density of states and group velocity between the two materials. The transmission coefficients for each side of the interface and for the other polarizations are given in the Supplementary Information.   

Our measured transmission coefficient profile thus indicates that low frequency, long wavelength phonons are transmitted to the maximum extent allowed by the principle of detailed balance, while high frequency, short wavelength phonons are nearly completely reflected at the interface. This result is consistent with many prior works in the literature. First, our result qualitatively reproduces the physically expected behavior that transmission coefficient increases with decreasing phonon frequency because phonons with sufficiently long wavelength do not perceive the atomistic disorder at the interface. The absence of acoustic reflections in our TDTR signal confirm the good acoustic match between the two materials and support our finding that long-wavelength phonons are mostly transmitted at the interface. Second, the measured transmission coefficient profile agrees with the experimental studies of polycrystalline silicon by Wang \emph{et al},\cite{Wang2011} which suggested that transmission coefficient should decrease with increasing frequency. Finally, the measurement also agrees with numerous molecular dynamics and atomic Green's function calculations, essentially all of which predict a decreasing transmission coefficient with increasing phonon frequency.\cite{Li2012b,Tian2014,Huang2010,Hopkins2009} Our work is thus able to provide direct experimental confirmation of these predictions for the first time while eliminating other possibilities for the transmission coefficients that appear in the literature.

\begin{figure*}[t!]
\centering
\includegraphics[scale = 0.6]{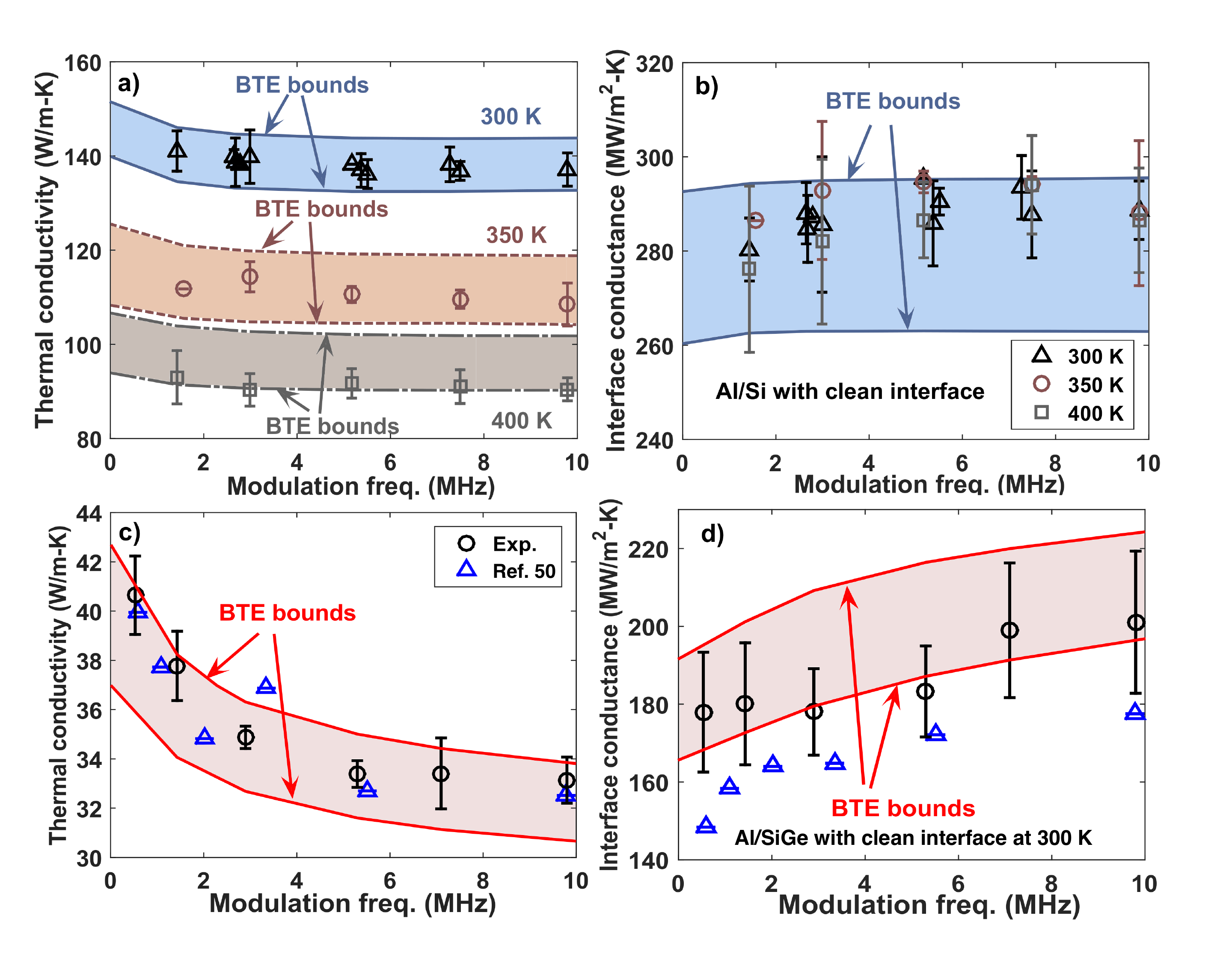}
\caption{\textbf{TDTR measurements on Al/Si  at different temperatures and Al/SiGe.} (a) Apparent thermal conductivity and (b) apparent interface conductance versus modulation frequency from experiments (symbols) and simulations (shaded regions) for Al on Si with a clean interface at 300 K, 350 K and 400 K. (c) Apparent thermal conductivity and (d) apparent interface conductance versus modulation frequency from experiments (circles: this work; triangle: Ref. 50) and simulations (shaded regions) for Al on SiGe with a clean interface. The magnitude and trend of the experimental data are reproduced using the same transmission coefficient profile as in Fig.~\ref{fig2}.}
\label{fig3}
\end{figure*}

Using this transmission coefficient profile, we plot the spectral interfacial heat flux versus phonon frequency and accumulative heat flux versus phonon wavelength in Figs.~\ref{fig2}(c) \& (d), respectively. In contrast to the prediction by the constant transmission coefficient profile, our results show that most of interfacial heat flux is carried by low to mid frequency phonons, with the contribution from high frequencies strongly reduced due to their small transmission coefficients. This observation highlights the importance of our ab-initio phonon transport modeling approach, as prior works neglected the contribution of these low-frequency phonons to interfacial heat flux.\cite{Wilson2013} In fact, we are unable to explain the magnitude of the observed interface conductance without the contribution of phonons of frequency less than 3 THz. Similarly, we find that we can only explain the measurements using the exact phonon dispersion for Al computed from DFT; simple dispersion relations such as Debye model cannot explain the data because they underestimate the contribution of low frequency phonons to thermal transport. 

Since the energy transmission at the interfaces is considered elastic, the transmission coefficient profile in theory should be independent of temperature. To confirm the robustness of the measured transmission coefficients, we conducted experiments at 350 K and 400 K and compared the experimental results with the calculations using the same transmission coefficient profile measured at 300 K. As shown in Figs.~\ref{fig3} (a) \& (b), the transmission coefficient profile measured at 300 K yields excellent agreement between simulations and experiments at higher temperatures.

To further support our measurements, we additionally measure the transmission coefficients for Al on SiGe. While this material has an additional point defect scattering mechanism compared to pure Si, we expect the transmission coefficients to be nearly the same given that the host lattice is unchanged. Figs.~\ref{fig3}(c) \& (d) plot the apparent thermal conductivity and interface conductance, demonstrating that our BTE modeling with the exact transmission coefficient profile shown in Fig.~\ref{fig2}(a) agrees well with the measured apparent thermal conductivity and interface conductance for Al/SiGe. This result confirms that the measured transmission coefficients for Si and SiGe substrates are indeed the same.  Importantly, the apparent thermal conductivity of SiGe differs from its actual thermal conductivity, measured by Thermtest as $50.7 \pm 0.5$ W/m-K using the transient plane source method on a bulk sample. This discrepancy highlights the strong influence of the transmission coefficient profile on the measurements from TDTR.

\begin{figure*}[t!]
\centering
\includegraphics[scale = 0.3]{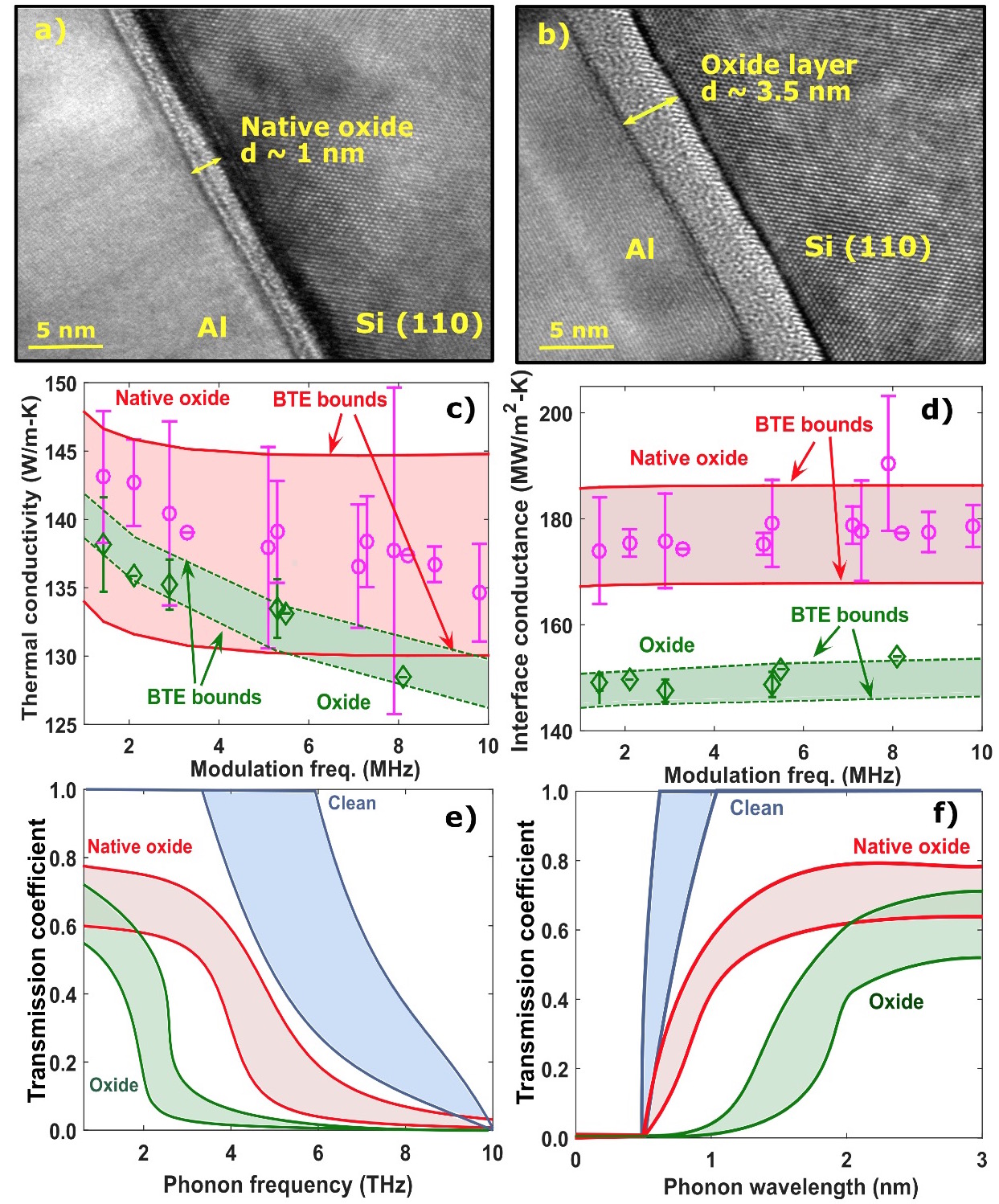}
\caption{\textbf{Transmission coefficient measurements of Al/Si with different types of interfaces.} TEM images showing the Al/Si sample with (a) native oxide layer (thickness $\sim 1$ nm) and (b) thermally grown oxide layer (thickness $\sim 3.5$ nm). (c) Apparent thermal conductivity versus modulation frequency and (d) apparent interface conductance versus modulation frequency of experiments (symbols) and simulations (shaded regions) for Al on Si with native oxide layer and Si with thermally grown oxide layer. The lines give the upper and lower bounds of the BTE simulations used to determine the transmission coefficients. The corresponding transmission coefficient profiles versus (e) phonon frequency and (f) phonon wavelength show that as the interface gets rougher, low frequency/long wavelength phonons are more likely to be reflected.}
\label{fig4}
\end{figure*}

So far, the measurements indicate that low frequency phonons are nearly completely transmitted through a clean interface. We next seek to determine what types of interfacial disorder can reflect these modes. We conducted additional measurements for Al on Si with a native oxide layer (thickness $\sim$ 1 nm as shown in a TEM image in Fig.~\ref{fig4}(a)) and Si with thermally grown oxide layer (thickness $\sim$ 3.5 nm as shown in a TEM image in Fig.~\ref{fig4}(b)). Since the oxide layers are sufficiently thin to neglect their thermal capacitance, we can treat them as part of the interface that modifies the transmission coefficient profile in our current BTE model. 

By again fitting the BTE results to the measurement as in Figs.~\ref{fig4}(c) \& (d), we are able to find the transmission coefficient profiles for these two cases as shown in Figs.~\ref{fig4}(e) \& (f). Compared to a clean interface, the transmission coefficients for Al on Si with a native oxide are reduced for all phonon modes. However, low-frequency phonons experience a larger reduction in transmission coefficient than do high-frequency phonons, which have transmission coefficients close to zero even at a clean interface. When the roughness of the interface increases with a thicker oxide layer, the transmission coefficient keeps decreasing and more phonons, especially those with wavelengths between 1 and 2 nm, are reflected at the interface.  Therefore, our measurements show that phonons with wavelength shorter than the interface roughness are more likely to be reflected by the interface than long-wavelength phonons, and as the interface gets rougher, a larger fraction of the phonon spectrum is affected by the interface.

\section*{Discussion}
Our work has considerable implications for thermal metrology and technological applications. First, TDTR is a widely used characterization tool to measure the thermal properties of thin films and bulk materials. However, this work demonstrates that in the quasiballistic regime, the apparent thermal conductivity and interface conductance measured by TDTR may not reflect bulk properties because both are strongly affected by the transmission coefficient profile. The apparently correct measurement of silicon thermal conductivity appears to be a fortuitous cancellation of two factors: the high transmission coefficient of low frequency phonons leads to an increased contribution to heat flux that offsets the deviation from Fourier's law that occurs due to a lack of scattering. If these two factors were not balanced, the apparent thermal conductivity of Si would not coincide with the bulk value. In fact, these factors are not balanced in SiGe, resulting in an apparent thermal conductivity that is substantially smaller than the actual value. Additionally, the apparent thermal interface conductance measured in SiGe deviates from the value measured in Si despite the fact that the transmission coefficients are exactly the same. These observations are further evidence that the apparent thermal properties extracted in a conventional TDTR interpretation approach do not necessarily reflect the actual physical properties of the materials. Although this situation is undesirable from a thermal metrology perspective, our work shows that TDTR is capable of providing considerable microscopic detail about thermal phonons provided the measurements can be properly interpreted. 

Second, our measurements show that the spectral profile of transmission coefficients is essential to understanding thermal transport across interfaces. Due to a lack of knowledge about interfaces, the phonon transmission coefficients are often taken to be a constant or calculated using the DMM. However, this work shows that both of these simple models are incapable of explaining the experimental measurements. Although the gray model and measured transmission coefficient profiles yield the same apparent interface conductance, the spectrum of heat transmitted across the interface is completely different, resulting in opposite conclusions regarding the primary heat carriers across the interface. Therefore, considering the spectral transmission coefficient profile across an interface is essential to accurately describing thermal phonon transport. 

Third, our work provides strong evidence that elastic transmission of phonons across an interface is the dominant energy transmission mechanism for materials with similar phonon frequencies. Our BTE model does not incorporate electrons or inelastic scattering yet is able to explain all of the measurements we performed. We conclude that inelastic transmission and coupling between electrons in metals and phonons in semiconductors have little influence on the energy transport for the materials considered here.  

Fourth, our results demonstrate that disorder at interfaces plays an important role in the spectral content of the heat transmitted through the interface and provides strategies to alter interface conductance. For instance, in applications like LEDs where the heat dissipation rate across interfaces is to be enhanced, the key to increasing interface conductance is to minimize the reflection of high frequency phonons by reducing defects; low frequency phonons are likely to be mostly transmitted already. On the other hand, the strong frequency dependence of the transmission coefficients can be exploited to create thermal phonon filters to selectively remove parts of phonon spectrum, analogous to optical long-pass filters. Phonons with wavelength much longer than the characteristic roughness of an interface are more likely transmitted through the interface while short-wavelength phonons are mostly reflected. The thermal phonon spectrum responsible for heat conduction can thus be manipulated by controlling the atomistic roughness of an interface. 

Finally, our work exemplifies the powerful insights that can be obtained through ab-initio phonon transport modeling, free of artificial parameters or arbitrary divisions of the phonon spectrum into diffusive and ballistic modes. Most of the previous studies focused on developing simplified thermal models to explain the modulation-dependent measurements for alloys such as SiGe but did not consider the contradictory observations in silicon. Obtaining the transmission coefficients from measurements that appear to be in good agreement with diffusion theory would be challenging without the predictive calculations reported here. Our work thus demonstrates a general approach that can be used to advance our knowledge of thermal phonons to a level on par with that of photons.

\section*{Methods}

\textbf{Sample preparation} Commercial high-purity natural Si (100) wafer and Si-Ge (1.5-2 at \% Ge) wafer (100) from MTI Corp. were used in the experiments. Before coating Al on the samples, three different surface conditions of the samples were prepared. First, the native oxide was removed with buffered HF acid to obtain a clean surface of Si and SiGe. After etching, the samples were immediately put into a vacuum chamber for Al deposition. Second, the native SiO$_2$ layer was left in place. No further treatment was taken for this condition before Al deposition. Finally, a thermally grown SiO$_2$ layer as fabricated by putting the Si samples into a tube furnace for three hours. The thickness of the native SiO$_2$ layer and thermally grown SiO$_2$ layer was measured by ellipsometry and TEM to be $\sim 1$ nm and $\sim 3.5$ nm, respectively. A thin film of Al was deposited on all samples using electron beam evaporator. The thickness of the Al transducer layer was 70 nm, measured by atomic force microscopy.

\textbf{TDTR measurements} The measurements are taken on two-tint TDTR. The details are available in Ref.~64.\nocite{Kang2008} The probe diameter is 10 $\mu$m and the pump diameter is 60 $\mu$m. Both beam sizes are measured using a home-built two-axis knife-edge beam profiler. With 60 $\mu$m pump heating size, the heat transfer problem can be treated as one-dimensional. All the measurements at $T = 300$ K are performed under ambient conditions, and the additional measurements at $T = 350$ and $400$ K are performed in an optical cryostat (JANIS ST-100) under high vacuum of $10^{-6}$ torr. 

\textbf{TEM images} The TEM samples were prepared by standard FIB lift-out technique in the dual beam FE-SEM/FIB (FEI Nova 600). To protect the top surface, a Pt layer with thickness $\sim$ 300 nm was deposited with electron beam evaporation followed by another Pt layer with thickness $\sim$ 3-4 $\mu$m by Ga ion beam. The lamella was cut parallel to the chip edge which was aligned to the wafer flat edge during initial cutting in TDTR sample preparation. As a result, the cutting surface normal was along (110) direction and all the TEM images were taken parallel to the Si (110) crystallographic zone axis. High resolution transmission electron microscopy (HRTEM) analyses were carried out in a FEI Tecnai TF-20 TEM/STEM at 200 kV. To avoid damage from the high energy electron beam, the beam exposure on region of interest was minimized especially at high magnification during operation.

\textbf{Ab-initio modeling} Thermal transport in TDTR experiments, assuming only cross-plane heat conduction, is described by the spectral, one-dimensional (1D) BTE under RTA\cite{Majumdar1993}. Efficiently solving this equation for the conditions of the TDTR experiment required two of our recent computational advances. First, we have obtained an analytical solution of the spectral BTE in a semi-infinite substrate subject to an arbitrary heating profile.\cite{Hua2014b} Second, we have employed a series expansion method to efficiently solve the spectral 1D BTE in a finite layer, suitable for the transducer film.\cite{Hua2014c} In this work, these two solutions are combined using a spectral interface condition\cite{Minnich2011b} that expresses the conservation of heat flux at each phonon frequency. Following the conclusions of prior computational works\cite{Lyeo2006,Murakami2014} that the energy transmission at the interfaces considered here is elastic, we enforce that phonons maintain their frequency as they transmit through the interface; direct electron-phonon coupling and inelastic scattering are neglected. The details of the calculation are in the Supplementary Information. The only inputs to our calculation are the phonon dispersions and lifetimes, calculated using DFT with no adjustable parameters by Jes$\acute{\text{u}}$s Carrete and Natalio Mingo, and the only unknown parameters are the spectral transmission coefficients.

\section*{Acknowledgements}

The authors thank J. Carrete and N. Mingo for providing the first-principles calculations for silicon, Prof. Nathan Lewis group for the access to the ellipsometer, and the Kavli Nanoscience Institute (KNI) at Caltech for the availability of critical cleanroom facilities. X. C. thanks Melissa A. Melendes, Matthew H. Sullivan and Carol M. Garland from the KNI for fabrication assistance, and Victoria W. Dix from the Lewis group at Caltech for the help with the ellipsometer measurements. This work was sponsored in part by the National Science Foundation under Grant no. CBET 1254213, and by Boeing under the Boeing-Caltech Strategic Research \& Development Relationship Agreement.

\clearpage

\bibliographystyle{is-unsrt}
\bibliography{FinalRef}

\end{document}